\documentstyle[twocolumn]{article}
\begin{document}
\openup6pt

\title {Do Scalar Tensor Cosmologies naturally explain all the current cosmological observations ? }
\author{A. Bhadra\thanks{ Email address: aru\_bhadra@yahoo.com}\\ 
High Energy and Cosmic Ray Research Centre\\  
University of North Bengal \\
Darjeeling (W.B.) 734 430 \\
INDIA\\
and\\
K.K. Nandi\thanks{ e-mail: kamalnandi@hotmail.com} \\
Department of Mathematics\\
University of North Bengal \\
Darjeeling (W.B.) 734 430 \\
INDIA
}
\date{}
\maketitle

\begin{abstract}
{\it It is shown that the general scalar tensor cosmologies may explain all the current cosmological observations without the need of invoking any ad hoc missing energy density. The explanation is based entirely on the internal dynamics of the theories. Two important predictions of the present analysis are: the universe is tending towards a matter dominated state with the dimunition of the dark energy component and the acceleration of the universe, if any, is slowing down with time. 
} 
\end{abstract}

Observations of small scale cosmic microwave background anisotropy on sub-degree angular scales [1-3] together with the cluster measurements [4,5]  and the luminosity-redshift relation of the type Ia supernova up to about $z \sim1$ [6] strongly indicate that  a significant component of the energy density of the universe has negative pressure [5,7].  Cosmological constant ($\Lambda$) is a straightforward and natural candidate for such a component [6]. However, models with fixed $\Lambda$ run into some serious difficulties such as the so-called ``cosmic coincidence problem'' (why $\Omega_{m}\sim \Omega_{\Lambda}$ right today) [5,8]. Moreover, the observational upper limit on $\Lambda$ is more than 120 orders smaller than what is expected naturally from a vacuum energy originating at the Planck time (the well known ``cosmological constant problem'') [9]. One is then led to models of dynamical vacuum energy or quintessence [10], involving a minimally coupled scalar field $\phi$ with a specific potential $U(\phi)$. Still the ``fine tuning problem'' remains; parameters in the scalar effective potential require some degree of fine tunings to avoid the coincidence problem [8]. Also, possible general couplings of $\phi$ with ordinary matter leads to spurious long range forces, violation of equivalence principle and the time dependence of gauge and gravitational constants [11]. To overcome the fine tuning problem, ``$k$-essence'' model is introduced [12] in which attractor dynamics at the onset of matter domination drives the scalar field into negative pressure state. Such a feature, however, comes at the cost of introducing a series of non-linear kinetic energy terms in the Lagrangian. There are also attempts to understand the scenario within the framework of non-Einsteinian theories of gravity. For instance, the observed acceleration of the universe could be modeled within the framework of general scalar tensor theories [13] admitting a potential term $U(\phi)$. The authors [14] underlined how the structure of the theory can be computationally determined from the observed luminosity distance and the linear density perturbations of a dust like universe. \\
The purpose of this letter is to show that the general scalar tensor cosmologies can naturally account for the matter density budget of the universe without the need of invoking any {\it ad hoc} missing energy density (in the form of a scalar potential or cosmological constant) and also explain the present accelerating expansion of the universe. To arrive at this conclusion, we use two natural inputs: the first one is that during cosmological evolution, the Einstein frame coupling function $\alpha(\tilde{\phi})$ ($\alpha^{2}(\tilde{\phi})  \equiv \frac{1} {2\omega(\phi) +3} $) tends toward its minimum (zero) or equivalently $\omega(\phi)$  approaches $\infty$ [15], where $\omega(\phi)$ represents the strength of the coupling between the scalar field $(\phi)$ and the curvature in the Jordan frame. This limit will be employed in the Jordan frame where $\omega(\phi)$ appears explicitly in the action. It should be of interest to note that the question whether the Jordan frame vacuum Brans-Dicke theory [16] ($\omega=$constant) reduces to general relativity in the limit $\omega \rightarrow \infty$ is being revisited in the current literature [17]. The other input is that the present value of  $\frac{\omega^{\prime}(\phi)}{\omega^{2} (\phi)} $ could be small but non-zero, where $\omega^{\prime} (\phi) \equiv \frac{d\omega (\phi)}{d \phi}$. Both these inputs are consistent with the constraints imposed on the scalar-tensor theories from local interactions. At present, observational limits from the solar system measurements are $\vert \omega(\phi) \vert > 3000$ and simultaneously $\vert \frac{\omega^{\prime}(\phi)}{\omega^{3} (\phi)} \vert  < 0.0006$ [18].  
One must also consider the cosmological constraints. The Big Bang nucleosynthesis in presence of a scalar field has been studied in several works [19, 20]. It imposes a stringent limit [20] on the post-Newtonian parameters $\gamma$ and $\beta$ to avoid over- or underproduction of $^4$He which  translates into a stronger bound on $\omega(\phi)$. Such a restriction is also automatically satisfied with the two inputs though, since we are studying only late time behavior of scalar tensor theories, the constraint due to nucleosynthesis is not directly applicable here.  \\
Since we perform our analysis within the framework of general scalar tensor theories, the weak equivalence principle, conservation laws and the constancy of the non-gravitational constants are automatically preserved. These theories are also found compatible with local situations such as the solar system experiments or binary pulsar tests [18, 21].  Moreover, in the extended inflationary scenario of cosmology, these theories admit natural termination of the inflationary era through the nucleation of bubbles without the need of finely tuned cosmological parameters [22]. We consider the most general scalar tensor theories of gravity with a single scalar field. In the Jordan conformal frame (since experimentally observed quantities are those that are written in the Jordan frame [15], we shall work in this frame throughout the paper) the general form of the action containing a massless scalar field $\phi$ is [13]
\begin{equation}
{\cal A} = \frac{1}{16 \pi }\int\sqrt{-g} d^{4}x \left[\phi R-\frac{\omega(\phi)}{\phi} g^{\mu\nu} \nabla_{\mu}\phi \nabla_{\nu}\phi +16\pi{\cal L}_{m} \right ]       
\end{equation}
where $R$ is the Ricci scalar constructed from the metric $g_{\mu\nu}$, and $ {\cal L} _{m} $ is the Lagrangian density of ordinary matter which could include electromagnetic field, nuclear field etc. The principle of equivalence is respected by requiring that the matter field Lagrangian can depend only on the metric $g_{\mu\nu}$ but not on $\phi$. \\
For a Friedmann-Robertson-Walker (FRW) universe with $ ds^{2} = -dt^{2} + a^{2}(t)\left[\frac{dr^2} {1-kr^{2}}+r^2d\Omega^{2} \right] $ and under dust approximation ($p=0$, $\rho=\rho_{o}\frac{a_{o}^{3}}{a^{3}}$), in the $\omega(\phi) \rightarrow \infty$ limit, the wave equation gives
\begin{equation}
\dot{\phi}\rightarrow \frac{A_{b}}{a^{3}\sqrt{(3+2\omega(\phi))}}
\end{equation}
where $A_{b}$ is a constant and expected to be negative. The Ricci scalar, in the limit $\omega(\phi) \rightarrow \infty$, is given by
\begin{equation}
R= -\frac{8\pi T}{\phi}-\frac{1}{2\phi}\left[\frac{3\omega^{\prime}(\phi)}{2\omega(\phi)^{2}} -\frac{1}{\phi }\right]\frac{A_{b}^{2}}{a^{6}} \;. 
\end{equation}
For compatibility with the weak field observations, a natural (and usual) choice is  $\frac{\omega^{\prime}(\phi)}{\omega^{2}(\phi)}\rightarrow 0$ [23]. However, present local observational constraints also admit a small non-zero value of $\frac{\omega^{\prime}(\phi)}{\omega^{2} (\phi)} $ which may arise from several situations such as: i) the evolution of the scalar field could drive $\frac{\omega^{\prime}(\phi)}{\omega^{2} (\phi)} $ toward a small non-zero positive value for a reasonable class of scalar tensor models. ii)  $\frac{\omega^{\prime}(\phi)}{\omega^{2} (\phi)} $ might be evolving towards zero but {\it at present} it has some small non-zero value. iii) the functional form of $\omega(\phi)$ could be chosen in such a way that $\frac{\omega^{\prime}(\phi)}{\omega^{2} (\phi)} $ is a non-zero constant (independent of $\phi$). This occurs, for instance, in the Einstein conformally coupled scalar field theory which can be rewritten in the form a scalar tensor theory with a specific form of $\omega(\phi)$ [24]. Hence, preserving the generality, we define $\frac{\omega^{\prime}(\phi)}{\omega^{2} (\phi)} \equiv \frac {\xi(\phi)}{\phi}$, where $\xi(\phi)$ is an arbitrary function of $\phi$. As $\omega(\phi) \rightarrow \infty$, the scalar field $\phi$ approaches a constant value ($\frac{1}{G}$) and so does $\xi(\phi)$. Currently we have only a rather weak observational limit $\vert \xi_{o} \vert < 1.8 G^{-1}$ (subscript o denotes present value throughout the paper). Though the evolution of the scalar field does not necessarily guarantee that this limit on $\xi_{o}(\phi)$ will hold simultaneously with $\omega(\phi) \rightarrow \infty$, such a condition is essential if the solar system tests are to accord with observations.
Using the expression of $\dot{\phi}$, the dynamical FRW equations under the limit $\omega(\phi) \rightarrow \infty$ become
\begin{equation}
\frac{\dot{a}^{2}+k}{a^{2}}=\frac{8\pi G \rho}{3}+\frac{A_{b}^{2}G^{2}}{12a^{6}} \;,
\end{equation} 
and 
\begin{equation}
2\frac{\ddot{a}}{a}+\frac{\dot{a}^{2}+k}{a^{2}}= -\frac{A_{b}^{2}G^{2}}{4a^{6}}(1-\xi(\phi)) \;.
\end{equation}
These are the master equations describing the dynamics of scalar tensor theories at late times. It is evident from these equations that, in the limit $\omega(\phi) \rightarrow \infty$, though scalar field approaches a constant value, there is a contribution from the scalar field (which is remarkably non-zero). This can be easily identified by comparing equations (4) and (5) with the corresponding equations of general relativity. However, in the Jordan frame it should not be regarded as the energy density of the scalar field rather it is a consequence of dynamical description of the gravitational field [25]. This gives rise to kind of ``matter field'' having a dynamical energy density $\frac{A_{b}^{2}G}{32 \pi a^{6}}$ and a pressure $\frac{A_{b}^{2}G}{32 \pi a^{6}}(1-\xi(\phi))$. This effective ``field'' has two components; one arises from the evolution of $\dot{\phi}$ with an equation-of-state similar to stiff matter ($p=\rho$) while the second component  results from evolving $\dot{\omega}(\phi)$ and contributes only to the pressure component. The equation-of-state of the effective ``field'' is $w \equiv p/ \rho_{eff} = 1-\xi(\phi)$ which may vary with time. For quintessence models also, $w_{q}$ varies with time but finally it tends to the equation-of-state for fixed cosmological constant $w_{\Lambda}= -1$. \\
The energy density of the effective field scales with respect to that of the background according to the law
\begin{equation}
\frac{\rho_{eff}}{\rho} \sim \frac{1}{a^{3}}
\end{equation}
which indicates that the universe has been tending towards a state of matter dominance ($\rho>> \rho_{eff}$). Note that in the vacuum energy (both static and dynamic) models, the reverse is true. Also it should be noted that Eq.(6) is valid only in the late times ($\omega(\phi) \rightarrow \infty$). At earlier periods, the ratio $\frac{\rho_{eff}}{\rho}$ could evolve quite differently and should depend on $\omega(\phi)$ as in the case of Brans-Dicke theory.
It follows from Eq. (4), that the present energy density is given by
\begin{equation}
\rho_{o}=\frac{3H_{o}^{2}}{8\pi G}\left(1-\frac{A_{b}^{2}G^{2}}{12H_{o}^{2}a_{o}^{6}} \right) + \frac{3k}{8 \pi G a_{o}^{2}} \;.
\end{equation}
and the critical density thereby becomes 
\begin{equation}
\rho_{c}^{ST} =  \frac{3H_{o}^{2}}{8\pi G}\left(1-\frac{A_{b}^{2}G^{2}}{12H_{o}^{2}a_{o}^{6}} \right)
\end{equation}
which is smaller than the $\rho_{c}^{GR}$. The above equation also puts a constraint on the numerical value of $A_{b}$; $A_{b}$ must be less than $\frac{2\sqrt{3}H_{o}a_{o}^3}{G}$ (in flat universe) to keep $\rho_{o}$  positive. 
If we consider our universe to be flat, as suggested from cosmological observations and also predicted by the inflationary paradigm of cosmology, Eq. (7) can be used to fix $A_{b}$ from  $\rho_{o}\sim .3\frac{3H_{o}^{2}}{8\pi G}$, the observed value [4,5]. Then  $\vert A_{b}/a_{o}^{3} \vert$ takes the value $\sim 10^{-7}$ $Kgm^{-3}s^{-1}$. 
It is worth mentioning that in the radiation epoch, the expression for $\dot{\phi}$ as given in Eq. (2) continues to remain valid [26] but the value of the constant $A_{b}$ is unlikely to be the same both at the radiation and matter dominated era. One has also to take into consideration the constraints due to phase transition. \\
Expression for the deceleration parameter $q_{o}$ is obtained from Eqs. (4) and (5):
\begin{equation}
2q_{o}= \frac{\rho_{o}}{\rho_{c}^{GR}}+ \frac{A_{b}^{2}G^{2}}{12H_{o}^{2}a_{o}^{6}} (4-3\xi_{o}(\phi))
\end{equation}
Thus there is no unique relationship between $q_{o}$ and $\rho_{o}$. An accelerating expansion of the universe ($q_{0}<0$) will occur if $\xi_{o}>\frac{4}{3}+ \frac{32 \pi H_{o} a_{o}^{6} \rho_{o}} {3 A_{b}^{2}G}$. For $q_{o} \sim -1$ [6], $\xi_{o}$ have to be $\sim 2.4$. The equation-of-state of the dark energy component of the universe should be determined from the proposed SNAP satellite mission [27]. Then, the present value of $\xi(\phi)$ will be precisely known. But it has to be also consistent with the findings of the future local gravitational experiments, such as Gravity Probe B, POINTS and Mercury Relativity Satellite mission.   \\
If we assume that the time dependence of $H(t)$ and $a(t)$ are close to those predicted by general relativity i.e., if  $H a^{3}$ is increasing with time, an important prediction emerging from Eq. (9)  is that the acceleration of the universe is {\it decreasing} with time. This is again in contrast with the prediction of a fixed cosmological constant or quintessence models in which acceleration will ever continue.  \\
For flat universe (k=0) and when $\xi(\phi) \rightarrow 0$, Eqs.(7) and (8) have three non-trivial exact solutions for $a(t)$. The only solution for which a(t) is positive throughout the history of the universe turns out to be 
\begin{equation}
a(t)=(rt^{2}+ st)^{\frac{1}{3}}
\end{equation}
where $r=6\pi G \rho_{o}a_{o}^{3}$ and $s=-\frac{\sqrt{3}}{2}A_{b}G$. The Hubble parameter (H) is given by
\begin{equation}
H= \frac{2rt + s}{3(rt^{2}+ st)}
\end{equation}
When $t  \rightarrow \infty$, both the scale factor and the Hubble parameter tend to corresponding GR values. When   $\vert A_{b}/a_{o}^{3}\vert $ is $\sim 10^{-7}$ $Kgm^{-3}s^{-1}$, the ratio $r/s$ in Eq. (10) is $\sim 10^{-18}$ and $a(t_{o}) \sim 1$. In that case, the age of the universe will be less (by a factor of 2) than that predicted by the Einstein-FRW model. One important point is that since $q_{o}$ is defined at the present epoch and the supernovae sample used in [6] cover a wide range of redshift, the estimation of acceleration is always within the context of a model of its origin. It will be of interest to examine whether an exact model like that given in Eq. (10) could give acceleration from the supernovae luminosity-redshift data. \\
It has been shown in the foregoing that the scalar tensor cosmologies are compatible with the all the current cosmological observations. Several evolving mechanisms simultaneously take place in scalar tensor cosmologies which may be responsible for different observed features of the universe. The ``attractor'' behavior [15,28] of scalar tensor theories toward general relativity for local interactions results from the evolution of $\omega(\phi)$, the evolving $\dot{\phi}$ may give rise to the socalled ``dark energy'' of the universe whereas the evolution of $\dot{\omega}(\phi)$ is found responsible for the present acceleration of the universe.   
Before closing, let us summarize the main results of the present analysis: \\
1) In the limit $\omega(\phi) \rightarrow \infty$, the scalar tensor theory equations resemble general relativity equations with a ``matter'' field having an equation-of-state $p=(1-\xi(\phi)) \rho_{eff}$.  \\
2) The critical density in scalar tensor cosmology is less than that in general relativity which explains the current observations [4,5]. \\
3) The universe has been tending towards a state of matter dominance, the density of dark energy component diminishing faster than that of ordinary matter. \\
4) The present observed [6] acceleration of the universe can be explained. However, the acceleration itself is decreasing with time.\\
The last two results are in fact two important predictions of scalar tensor cosmologies. Future cosmological observations, like the proposed SNAP satellite mission and the MAC and the PLANCK cosmic microwave experiments, might have the potential to test these predictions. \\

{\bf \it Acknowledgment :}
AB wishes to thank IUCAA Reference Centre (NBU) for providing its facility and to G. Majumder (TIFR) for stimulating comments and various help.


\begin{thebibliography}{99}
\bibitem{ab:23}The limit $\frac{\omega^{\prime}(\phi)}{\omega^{2}(\phi)}\rightarrow 0$ or a finite value renders the Ricci scalar well-determined [26]. Furthermore, it is also consistent with the PPN constraint $\frac{\omega^{\prime}(\phi)}{\omega^{3}(\phi)}\rightarrow 0$ arising in the local problems.
\end{thebibliography}
\end{document}